\documentclass[showpacs,amsmath,amssymb,floatfix,prl,twocolumn]{revtex4}
\usepackage[pdftex]{graphicx}
\usepackage{amssymb,amsfonts,amsmath}
\usepackage{graphicx,amsmath}
\usepackage{float}
\usepackage{color}
\usepackage{siunitx}
\def\s{\sigma}

\begin{document}
\bibliographystyle{unsrt}
\title{Field-enhanced recombination at low temperatures in an organic photovoltaic blend}
\author{S.~Athanasopoulos}
%\email{as2257@cam.ac.uk}
%\thanks{}
\affiliation{Cavendish Laboratory, University of Cambridge, J. J. Thomson Avenue, Cambridge CB3 0HE, UK}
\author{N.~C.~Greenham}
%\email{ncg11@cam.ac.uk}
%\thanks{}
\affiliation{Cavendish Laboratory, University of Cambridge, J. J. Thomson Avenue, Cambridge CB3 0HE, UK}
\author{R.~H.~Friend}
%\email{rhf10@cam.ac.uk}
%\thanks{}
\affiliation{Cavendish Laboratory, University of Cambridge, J. J. Thomson Avenue, Cambridge CB3 0HE, UK}
\author{A.~D.~Chepelianskii}
%\email{chepelianskii@lps.u-psud.fr}
%\thanks{}
\affiliation{Cavendish Laboratory, University of Cambridge, J. J. Thomson Avenue, Cambridge CB3 0HE, UK\\
 LPS, Universit\'e Paris-Sud, CNRS, UMR 8502, F-91405, Orsay, France}

% 
% TODO for Alexei : 
% 2. Modify Figure 2 in the light of our discussion if appropriate
% 3. Decide what to do about the "dissipative" section - I'm sorry I have forgotten what we discussed (did we agree it was not vital to the argument?)
%


\begin{abstract}
We report on the non-trivial field dependence of charge carrier recombination in an organic blend at low temperatures. A new microwave resonance technique for monitoring charge recombination in organic semiconductors at low temperatures is applied in bulk heterojunction P3HT:PCBM blends with results showing that an external electric field can in fact increase recombination. Monte Carlo simulations suggest that this contradiction to conventional wisdom relates to electron-hole pairs that are separated at donor-acceptor interfaces where the electric field acts in synergy with their Coulomb attraction. For this behaviour to occur a critical initial separation of $\sim$ \SI{5}{\nm} between the carriers is required.
\end{abstract}

\maketitle

\section{Introduction}

The recombination of charges in disordered molecular semiconductors is key to the operation of organic electronic devices such as light-emitting diodes (LEDs)~\cite{KuikAdvMat14} and solar cells~\cite{DeibelRepProgPhys10}.  The bimolecular encounter of opposite charges is known as non-geminate recombination, and is an important loss mechanism limiting the photovoltage and photocurrent in organic solar cells~\cite{ShuttlePRB08,GirishAnRevPhysChem14,Nguyen13}, whereas in LEDs it is the process leading to the desired light emission.  In some organic solar cells geminate recombination, where electrons recombine with holes arising from the same photogenerated exciton, can also play an important role~\cite{MihailetchiPRL04}.

Theoretical models for recombination must account for the hopping of charges within a disordered density of states~\cite{BaranovskiiPhysStatSolB14}, in the presence of a Coulomb attraction between electron and hole that is long-range due to the low dielectric constants of organic semiconductors. The simplest approach is the Langevin model which predicts a non-geminate recombination rate, $\nu_r$ proportional to the product of the electron and hole carrier densities $n_{e}$ and $n_{h}$, with a prefactor that is proportional to the sum of their mobilities, giving $\nu_r \propto \gamma_{L} n_e n_h$, with $\gamma_L=e(\mu_h+\mu_e)/(\epsilon_0 \epsilon_r)$, where $\mu_e$ and $\mu_h$ are the electron and hole mobilities respectively, and $\epsilon_{r}$ is the relative permittivity of the material. This model has been widely applied but has some well-established problems~\cite{GirishAnRevPhysChem14,Nguyen13}, most notably the fact that in some photovoltaic systems the measured recombination is greatly suppressed compared with the predicted Langevin value~\cite{KosterAPL06,GrovesPRB08}.  More realistic models need to go beyond the effective medium approach and take into account the microscopic density of states, including any deep trap sites~\cite{KirchartzPRB11,KuikPRL11},  the possibility that carriers reaching short separations do not necessarily recombine rapidly and terminally to the ground state~\cite{PhilipJACS14}, and the presence of a nanostructure in bulk heterojunction photovoltaic materials that confines electron and hole to spatially distinct materials.  These effects are best addressed by Monte Carlo simulations of the microscopic recombination process~\cite{WatkinsNano05,MarshJAP07,GrovesJCP08,GrovesPRB08}, although these simulations lack the simplicity of analytical models.

From an experimental point of view it is not easy to  measure bimolecular recombination rate constants directly. Transient photovoltage measurements suggest that the rate constant increases strongly with carrier concentration~\cite{LiAFM11}, but  measuring this carrier concentration independently is not always straightforward. Transient absorption measurements also provide kinetic information~\cite{DhootChemPhysLet02,JennyPRB03}, but it can be difficult to distinguish between electrons and holes, and between free and trapped charge.  Recently, comparison of double-carrier and single-carrier devices have been used to infer Langevin prefactors~\cite{GaoAEM12}.

Low-temperature recombination behaviour has received much less attention~\cite{SchultzPRB01}. This is partly because most practical device applications occur at room temperature and partly because low temperatures are more likely to lead to the out-of-equilibrium transport regime of  photo-generated carriers that is difficult to tackle theoretically. However, these non-equilibrium effects are important to understand since in many materials disorder is high enough for them to play a role even at room temperature~\cite{HeinzTopCurChem12}.  This is particularly true for blends, where the nanostructure mimics the effects of positional and energetic disorder, and in the relatively thin films used for typical devices, where carriers may recombine or be extracted before reaching quasi-equilibrium~\cite{NikitenkoJPhysCondMat07}. Since the transport properties depend on the ratio of disorder to thermal energy~\cite{HeinzPhysStatSol93,BorsenbergerPhysStatSolA93,StavrosPRB09,HoffmannJPCC12}, low-temperature measurements provide a means to study the effects of disorder without varying the material.  Arkhipov has dealt with the problem of low temperature monomolecular and bimolecular recombination in the absence of an electric field using an analytical hopping model~\cite{ArkhipovJPhysCondMat} while Shklovskii has derived expressions for geminate pair radiative recombination at low temperature assuming diffusive hopping motion~\cite{ShklovskiiPRL89}.

Here we investigate the dependence of charge carrier recombination on external field in an organic blend at low temperatures. By monitoring the resonance frequency shift of a superconducting resonator~\cite{AlexeiPRL14} we show that an applied electric field can actually increase the recombination of long-lived carriers in bulk heterojunction P3HT:PCBM blends. To understand this behaviour, we have simulated the long-time dynamics of photo-generated carrier pairs with kinetic Monte Carlo simulations of hopping transport using energetically disordered lattice morphologies. We find that the increase in recombination with electric field is due to carrier pairs at donor-acceptor interfaces with an initial critical separation of at least \SI{5}{\nm}. This rather large initial separation length reveals a new recombination regime where the Coulomb interaction and the external electric field act synergistically to enhance recombination.

\section{Experimental method to probe the density of optically excited carriers}

At cryogenic temperatures photo-generated charges are trapped within nanometer-size domains 
due to the large energetic disorder in organic semiconductors, which is typically around \SI{100}{\meV}~\cite{HeinzPhysStatSol93}.
Localized carriers cannot contribute directly to the conductivity of the photovoltaic layer 
however they can manifest themselves through an increase in its dielectric polarizabiliy,
since each carrier can be displaced within its localization volume by the applied AC electric field.
Thus, the dielectric constant of a photovoltaic blend increases under illumination at low temperatures.
This effect can be detected by placing the sample in a microwave cavity and by monitoring 
the drop of the cavity resonance frequency $f$ as the sample capacitance increases under illumination \cite{AlexeiPRL14}.
Similar experiments have been performed at room temperature on triplet excitons in order to determine 
their polarizability volume \cite{WarmanSPIE,Warman99}. Room-temperature experiments were possible in that case because 
the triplet excitons retain a dielectric response even at high temperatures due to their large Coulomb binding energy.

The requirement for low temperatures to study the dielectric properties of photo-generated carriers enables 
the application of sensitive cryogenic detection techniques. 
At temperatures around a few Kelvin, a superconducting strip-line resonator offers several experimental advantages.
It can provide both a high quality factor and a good coupling between the electromagnetic modes of the resonator 
and the thin region of photoexcitation in a film (the optical absorption depth in organic semiconductors is usually around \SI{100}{\nm}).
While strip-line superconducting resonators have only recently been applied 
to study photoinduced charge transport \cite{AlexeiPRL14,BlueRay}, they are frequently used in low-temperature condensed matter physics as a highly sensitive detector.
For example strip-line superconducting resonators have been employed successfully for investigations of persistent currents in mesoscopic rings \cite{Reulet95,DeblockPRL02,DeblockPRB02},
and for dynamical studies in mesoscopic superconductivity \cite{Chiodi,Dassonneville}. They have also allowed a higher sensitivity to be achieved 
in spin-resonance experiments compared with techniques relying on conventional cavities \cite{Tyryshkin13,Tyryshkin14},
achieving the strong coupling limit with a small ensemble of spins \cite{Esteve}.

\begin{figure}
\centerline{\includegraphics[clip=true,width=8cm]{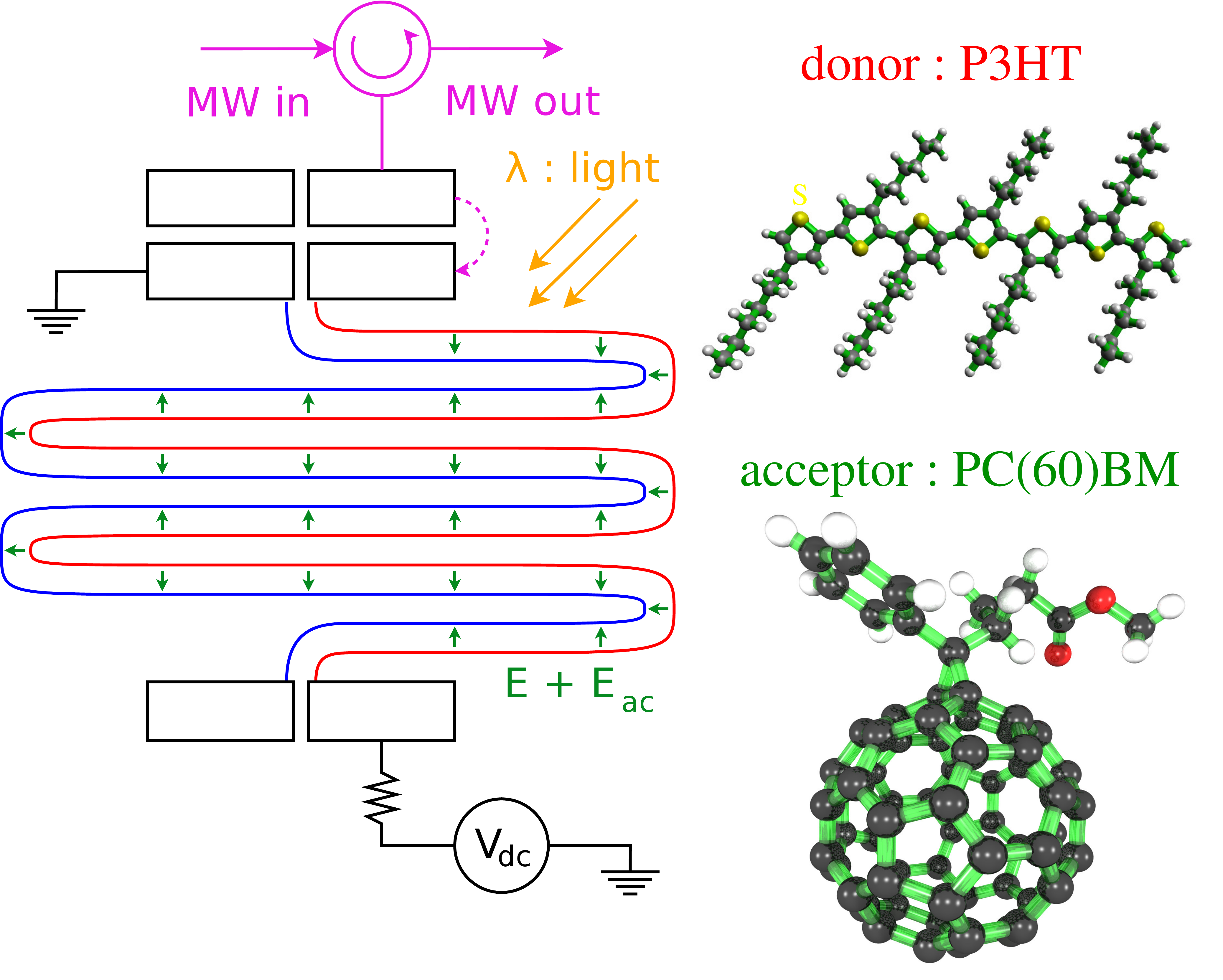}}
\caption{(Colour online) Left: Schematics of the experiment. Nb meanders are covered by a thin P3HT:PC(60)BM layer
which is coupled to the AC electric $E_{ac}$, in addition a static electric field $E$ can 
also be created by applying a static voltage $V_{dc}$ on one of the meanders through a coupling  
impedance.
The sample is optically excited at a tunable wavelength $\lambda$, typically $\lambda = \SI{550}{\nm}$.
Right: The chemical structure of the donor/acceptor molecules forming the P3HT:PC(60)BM layer.
}
\label{FigSetup}
\end{figure}

We recently investigated the photo-induced dielectric signal in P3HT:PC(60)BM blends spin-coated 
on a Nb superconducting resonator, showing that the amplitude of the dielectric signal was proportional to the concentration of photo-induced charges
at carrier concentrations $n_S$ below \SI{4e24}{\m^{-3}}.
This was confirmed by comparing the dielectric and the light-induced spin resonance signals, 
and allowed a quantitative conversion to be established between the relative shift of the resonator frequency 
and the density of photo-generated charges:
\begin{align}
\frac{\delta f}{f} = -\frac{n_S a_l^3 W}{\lambda_E} .
\label{dftone}
\end{align}
In this expression $n_S$ represents the photogenerated carrier density which is equal to twice the 
electron density due to the presence of an equal number of holes, and $a_l^3$ is the average polarization volume 
associated with the carriers (in P3HT:PC(60)BM we found $a_l = \SI{2.3}{\nm}$ \cite{AlexeiPRL14}).
The quantity $W \simeq \SI{130}{\nm}$ is the estimated PV layer thickness, and $\lambda_E =  \SI{5.6}{\micro\m}$ 
is the effective confinement length of the AC electric field in the resonator. 
We found that at a temperature of a few Kelvin, the mean recombination time $\tau_{rec}$ of the carriers diverged 
very steeply at low carrier density, giving rise to extremely long-lived carriers in the low-density limit. 
For example at $n_S = \SI{e24}{\m^{-3}}$ we found a mean recombination lifetime $\tau_{rec} \simeq \SI{e2}{\s}$
with a lifetime dependence on density which approximately followed a $\tau_{rec} \propto n_S^{-5}$ scaling.

In the experiments that we present here we introduce the possibility to apply a static electric field in addition to the AC electromagnetic field in the resonator, 
enabling  investigation of the effect of a static electric field on charge recombination at low temperatures. 
The  PV layer studied was a P3HT:PC(60)BM blend with a 1:1 weight ratio spin-coated from chlorobenzene and annealed at \SI{130}{\celsius}.
The resonator used in our experiments consists of two folded superconducting meanders.
A DC voltage source was connected to one of the meanders through 
an impedance of \SI{10}{\mega\ohm}, much larger than the characteristic impedance of the meander transmission line (see  Fig.~1). 
Under these conditions the microwaves are mainly reflected at the DC coupling impedance as a result of the large impedance mismatch,
which reduces the resonator quality factor by about a factor of two. The AC potential was coupled through a capacitor 
to the DC polarized meander, while the opposite meander was grounded. In this configuration the magnitude of the 
 electric field is given by the voltage bias divided by the spacing between the meanders, measuring $\sim \SI{5}{\micro\m}$.
Typical values of the electric field $E$ in our experiments were in the range $\SI{e6}{\volt \m^{-1}}-\SI{e7}{\volt \m^{-1}}$.
Illumination of the sample was provided by a xenon arc lamp coupled to the sample through an optical fiber.

\begin{figure}[h]
\centerline{\includegraphics[clip=true,width=8cm]{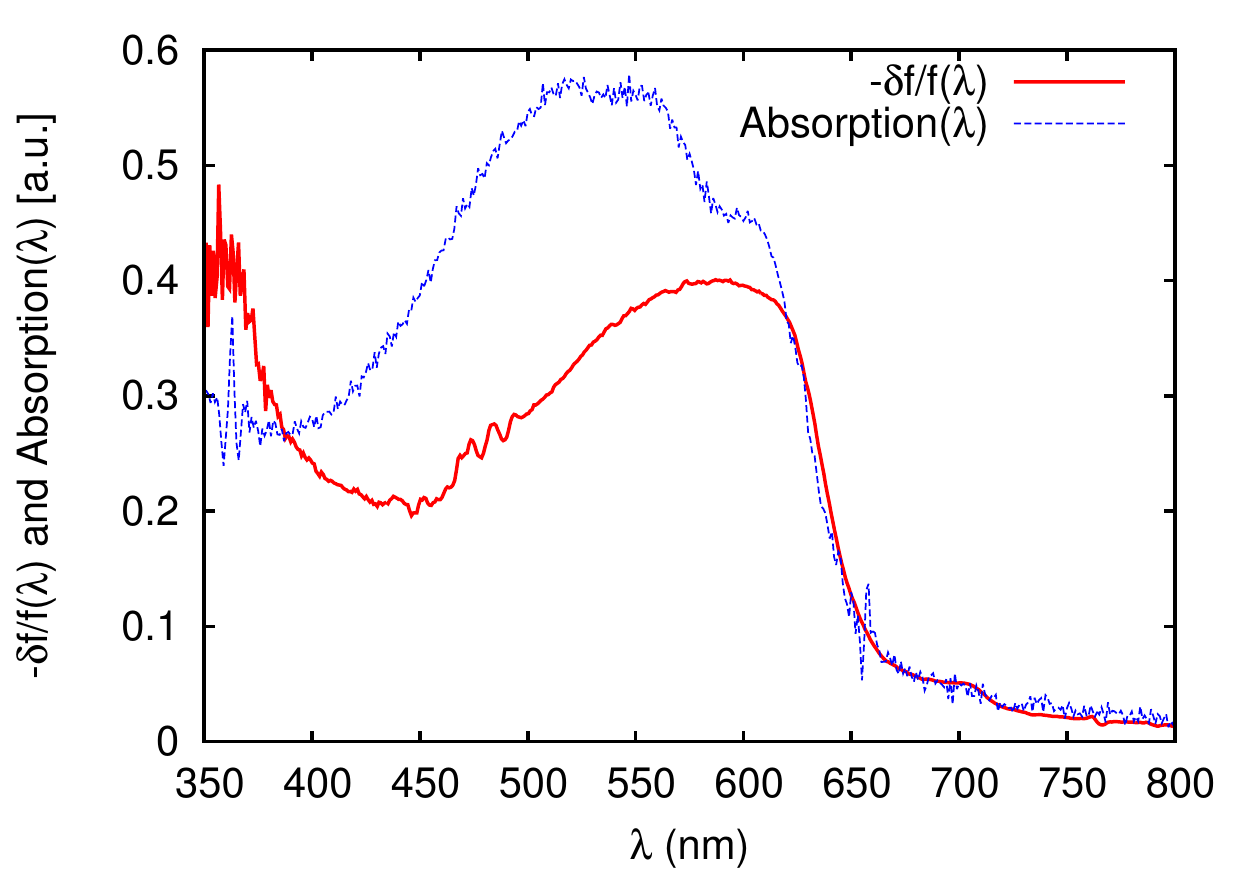}}
\caption{(Colour online) Dependence of the cavity frequency shift on the excitation wavelength 
measured using the double modulation technique described in the text.
The frequency shift is normalized by the intensity at wavelength $\lambda$ 
incident on the sample. 
The results are compared with the absorption coefficient measured on the same 
sample with an UV-Vis spectrometer. We note that the absorption edge at \SI{650}{\nm} coincides in the two 
experiments. The microwave measurements were performed at a temperature of \SI{4.2}{\kelvin} whereas the absorption measurements where performed in a UV-Vis spectrometer at room temperature.
}
\label{FigSpectrum2}
\end{figure}

To confirm that the frequency shift of the resonator was indeed due to photoexcitations in the polymer film,
we changed the illumination wavelength by filtering the white light generated by the xenon lamp with a monochromator.
This allowed us to compare the wavelength dependence of the signal from the resonator with the absorption spectrum
of the P3HT:PC(60)BM film which was measured on the same sample at room temperature. The experimental results, shown in Fig.~\ref{FigSpectrum2},
confirm the overlap between the two spectra and demonstrate that the resonator is indeed probing photo-excitations in the PV layer. 
We note that the agreement between the two measurements is only qualitative. A quantitative comparison would require a precise calibration 
of the number of absorbed photons in the PV sample as function of wavelength under the cryogenic experimental conditions of the microwave experiment.
This calibration is experimentally challenging and was not performed in our experiments. 

To enhance the detection sensitivity in this experiment we developed a double modulation technique 
where both the microwave and the light signal were modulated.
The microwave signal, with typical power around \SI{1}{\micro\watt}, was frequency modulated at $f_{FM} = \SI{317}{\Hz}$ and the power reflected from the resonator 
detected with with a lock-in amplifier referenced to $f_{FM}$. The output of this first demodulation stage was fed into a proportional/integral controller 
which determined the position of the resonance frequency based on the microwave reflection signal.
With our gain and time-constant settings the response time of the feedback circuit to an abrupt 
change of microwave carrier frequency was around \SI{50}{\milli\s}. 
The resonance frequency signal generated by the controller was then in turn demodulated against the light chopping frequency (typically \SI{7}{\Hz}) using a second lock-in amplifier. This allowed us to achieve a very high sensitivity to photo-induced changes 
in the resonator frequency with relative accuracy $\delta f/ f \simeq 10^{-9}$.
The  $\delta f$ obtained was then divided by the irradiation intensity at 
the wavelength used, measured at the output of the optical fiber,
thus compensating for wavelength dependence of the emission spectrum of the xenon lamp and monochromator throughput. In the first experiments below, only the first modulation stage at frequency $f_{FM}$ was used
and the chopper wheel giving the light modulation was removed from the experiment. 

\section{Effect of an electric field on the photo-generated charge density}

In the previous section we showed that the frequency shift of the superconducting resonator $\delta f$  under illumination 
is proportional to the density of photoinduced charges. The measurement of this frequency shift allows us to follow the carrier population 
as a function of time under an applied electric field $E$.
To investigate the effect of this electric field on the carrier density we adopted the following experimental procedure.
The resonator, initially held in the dark, was illuminated at time $t = 0$ with continuous \SI{550}{\nm} light from the monochromator. 
The illumination leads to a progressive build up of the charge population in the PV layer 
which increases $|\delta f|/f$. At a time $t = t_1$  
a static electric field $E$ is applied across the meanders until a later time $t_2$ when it is turned off again. 
We fixed the time interval $\Delta t = t_2 - t_1$ when the field is applied
equal to the duration of the initial charge accumulation time $\Delta t = t_1$.
We emphasize that the sample is continuously illuminated during the entire static electric field on/off sequence.
At the end of the recording the illumination is turned off and the sample is left 
in the dark for around an hour to allow relaxation of the photoexcited carriers. This rather long waiting time between the measurements is imposed by 
the very long charge recombination rates at low temperatures in the low-density regime~\cite{SchultzPRB01}.

\begin{figure}
\centerline{\includegraphics[clip=true,width=8cm]{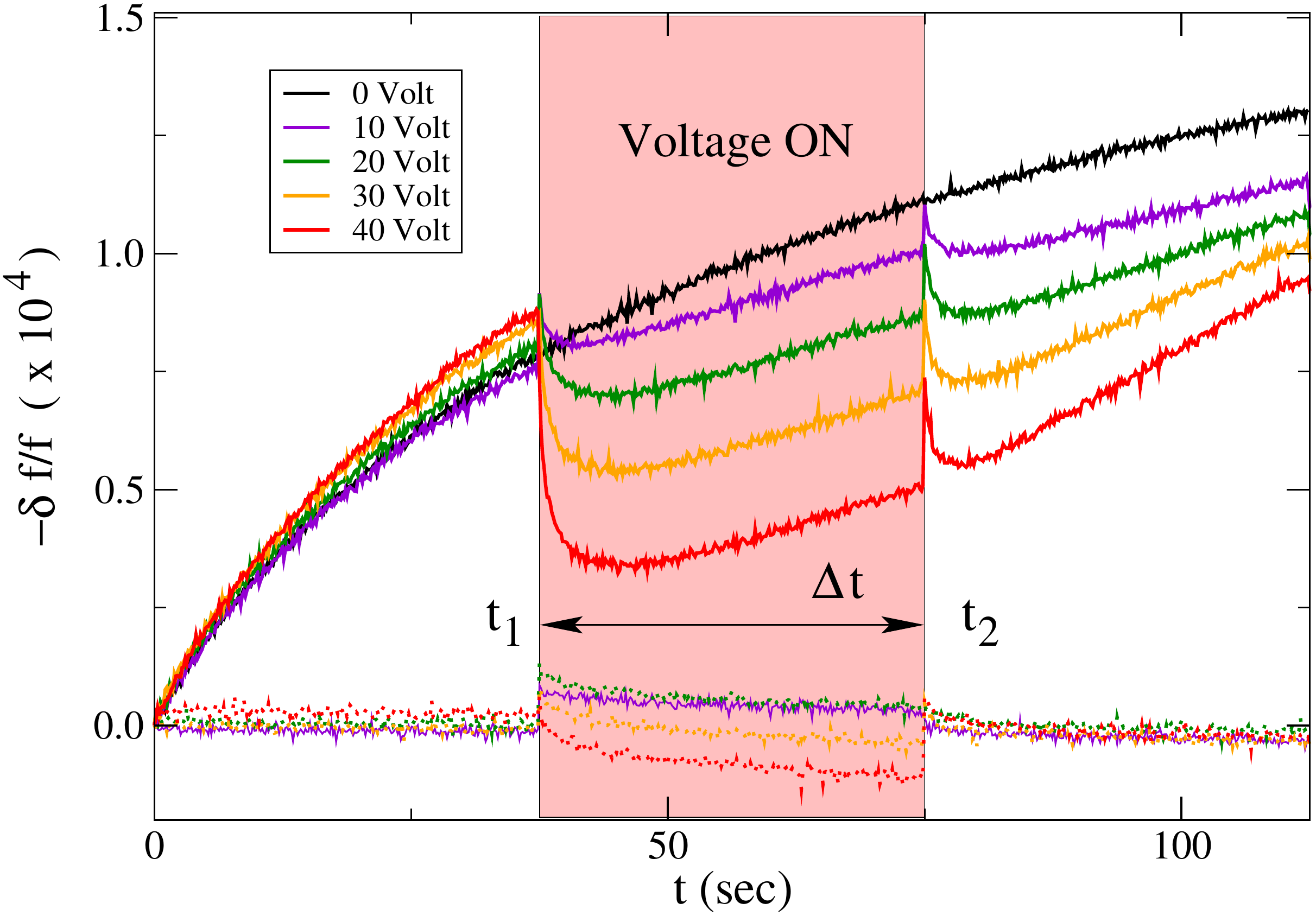}}
\caption{(Colour online) Relative change of the cavity resonance frequency as a function of time for different amplitudes of the  voltage applied across the meanders. This voltage was applied only between times $t_1$ and $t_2$ (for a total duration $\Delta t = t_2 - t_1$) and was set to zero outside this time interval. For the thick lines, the PV layer was continuously illuminated 
stating from time $t = 0$ with an estimated incident optical power of around \SI{10}{\nano\watt}, whereas for the dotted lines the sample was kept in the dark as a control experiment. 
}
\label{FigDeltaF}
\end{figure}

Typical experimental traces obtained during the sequence described above are shown in Fig.~\ref{FigDeltaF}. They all exhibit a pronounced drop of $|\delta f|/f$
when the static electric field is applied at time $t = t_1$. This drop is not recovered when the static field is removed at $t_2 = t_1 + \Delta t$,
demonstrating that the electric field leads to an irreversible reduction of the photoinduced dielectric signal.
This observation allows us to discriminate between field-induced recombination which is irreversible,
and the effect of the static electric field on the dielectric polarizability of the trapped carriers 
that would return to its original value after the removal of the static electric field. 
To rule out any significant effect of the DC bias on the resonator itself,
or on any left-over trapped carriers present at the start of the trace, we also performed control experiments where the DC voltage was applied to the resonator without illumination. The traces displayed in Fig.~\ref{FigDeltaF} confirm that the background effect is much weaker than 
the photoinduced signal in the DC bias range explored.
We consider that the main source of this background signal is related to the electrical cross-talk between 
the sensitive feedback circuit tracking the resonance frequency and the on/off voltage pulses applied to the meanders.
This illumination-independent signal was subtracted from the experimental traces.

\begin{figure}
  \centerline{\includegraphics[clip=true,width=8cm]{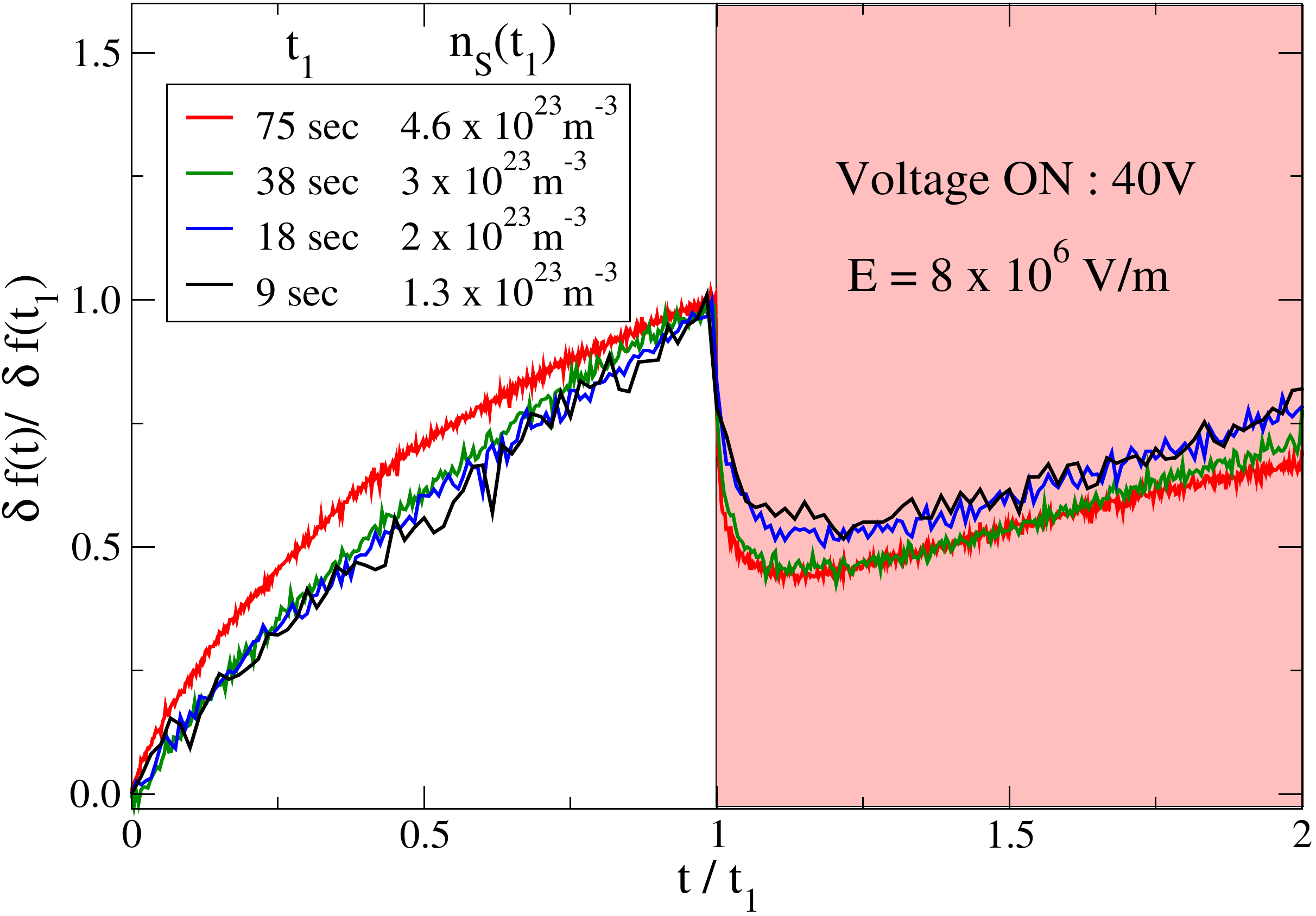}}
\caption{(Colour online) Relative change of the cavity resonance frequency as a function of time for different durations of the delay $t_1$ between the beginning of 
carrier accumulation at time $t = 0$ when the illumination starts and the application of the static bias.
The photoinduced carrier density at the time of the voltage onset, $n_S(t_1)$, increases with $t_1$. Its value for the different traces was determined from 
$\delta f(t_1)$ and is indicated on the legend.
Time is normalized by the carrier accumulation time $t_1$, and the frequency shift of the resonator $\delta f$ is normalized by its value at time $t_1$.
All the traces collapse onto the same dependence which does not depend on the photogenerated carrier density $n_S(t_1)$.
}
\label{ScaledDf}
\end{figure}

In order to understand how the observed reduction of the photo-polarizability signal depends on the 
carrier density we varied the duration $t_1$ of the initial charge accumulation stage.
Figure~\ref{ScaledDf} shows the photoinduced frequency shifts $\delta f(t)$ obtained for different charge accumulation times $t_1$.
They are rescaled by their value $\delta f(t_1)$ at time $t_1$ just before the static field is applied.
A longer charge accumulation time $t_1$ leads to a higher induced charge density $n_S(t_1)$ at the field onset time; 
the corresponding $n_S(t_1)$ value can be estimated from the frequency shift $\delta f(t_1)$ using Eq.~(\ref{dftone}).
This leads to values of $n_S(t_1)$ in the range \SI{1.3e23}{\m^{-3} } to \SI{4.3e23}{\m^{-3}},
as indicated in the figure. 
Even when the charge density was varied by more than a factor 3, 
the rescaled traces collapse accurately on a single curve. 
Thus, the drop in the carrier population induced by the electric field 
is proportional to the number of photogenerated carriers even in the low-density limit. 

\begin{figure}
  \centerline{\includegraphics[clip=true,width=7cm]{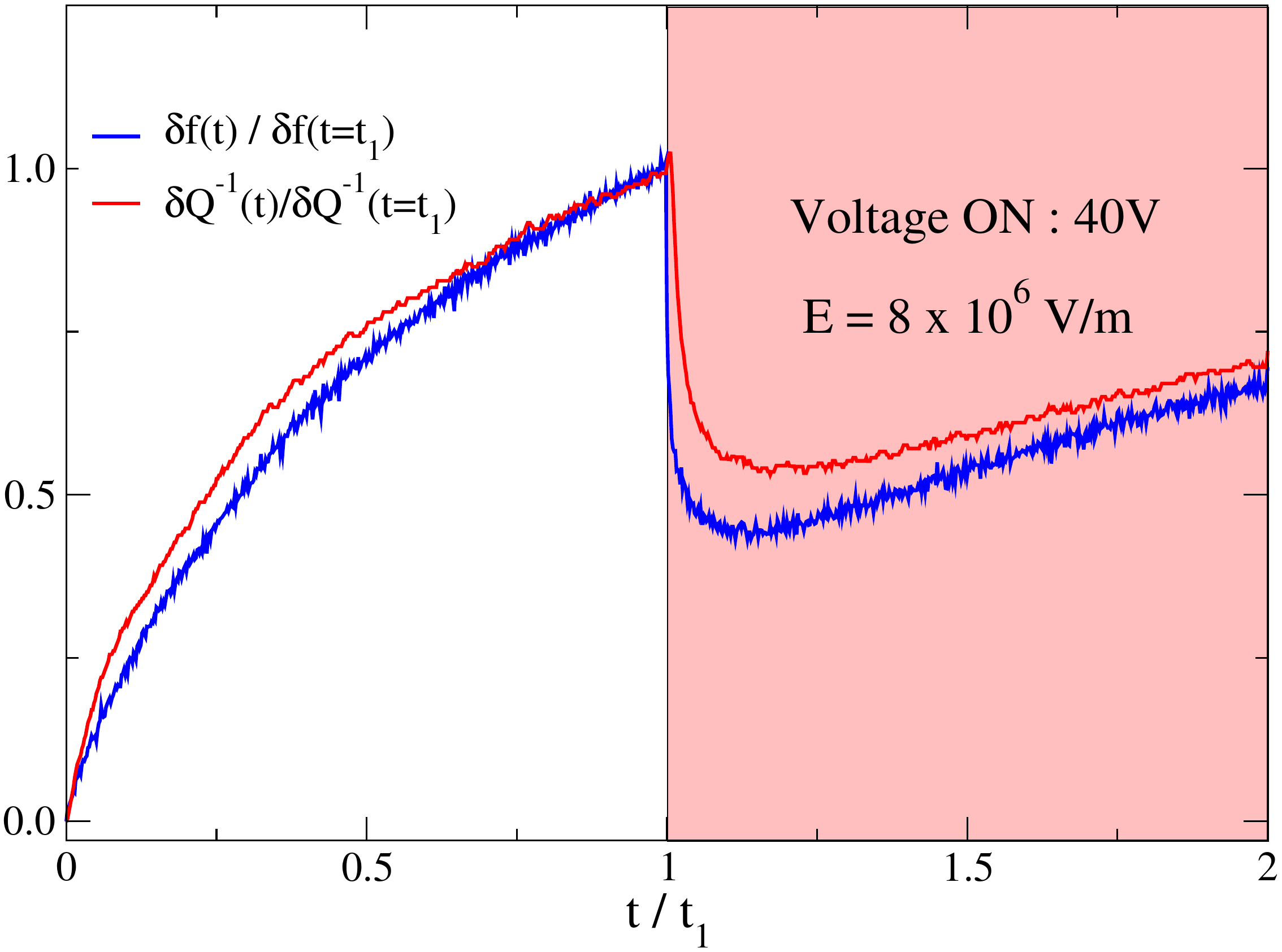}}
\caption{(Colour online) Comparison between the time evolution of the cavity losses and of the frequency shift when the static voltage is applied across the meanders with $\Delta t = 75\;{\rm s}$. 
The cavity losses are quantified by the change of the cavity inverse quality factor $\delta Q^{-1}$, the values of $|\delta f|$ and $\delta Q^{-1}$ are rescaled by their maximal value attained just before the static electric field is applied. 
}
\label{FigDQ}
\end{figure}

It is interesting to understand if the observed recombination occurs due to changes in the carrier dynamics on a fast timescale comparable with the microwave oscillation period or if it is a consequence of changes in the slower motion corresponding to hopping between sites. 
We can probe experimentally the dynamics on a fast time scale by measuring the dissipative response of the resonator.
Indeed carriers do not respond perfectly in phase to the AC driving field created by the resonator; 
instead their phase lags behind giving rise to a dissipative response, $\delta Q^{-1}$.
This dissipative response becomes important only when the frequency of the driving field 
becomes comparable with timescale on which the carriers can respond to a polarizing field within their trap states. The dielectric response of the carriers as a function of frequency is well described by a Debye response function 
and it is thus possible to introduce a characteristic timescale: the Debye time $\tau_D$. Its value can be obtained from the ratio of the dissipative response $\delta Q^{-1}$
to the dielectric response $\delta f$:
\begin{align}
\tau_D = 2 \pi \frac{\delta Q^{-1}}{\delta f}.
\end{align}
This value is characteristic of the trap states; its typical value for the P3HT:PCBM blend under our experimental conditions is approximately $\tau_D \simeq 18\;{\rm ps}$ \cite{AlexeiPRL14}.
To probe the behavior of this quantity under an applied electric field we measured the change in the cavity inverse quality factor $\delta Q^{-1}$ with illumination time.
This quantity was measured using the response at the second harmonic of the frequency modulation frequency $f_{FM}$ as described in \cite{Chiodi,Dassonneville}.
We find, as shown in the experimental traces Fig.~\ref{FigDQ}, that the time evolution of the cavity losses
coincides with the time evolution of the dielectric properties.  These results indicate that the Debye time is not significantly changed during the voltage on/off sequence, suggesting that the dominant effect of the electric field is on the slower hopping dynamics.

To summarize, the experiments presented show a substantial increase in the recombination rate of photogenerated charges 
at cryogenic temperatures due to the presence of a static electric field. We have also demonstrated that the field-induced signal is proportional 
to the density of photogenerated carriers even in the limit of low carrier density.
This suggests that geminate recombination is the main carrier loss mechanism involved as opposed to bimolecular recombination.

\section{Monte Carlo simulations of field-dependent low-temperature recombination}
To understand the influence of an electric field on the long-time dynamics of photogenerated carriers we have performed Monte Carlo simulations on lattice morphologies. The important ingredients for the model are the lattice morphology and the rates for transport.
We consider two type of morphologies, a bulk heterojunction blend and a bilayer. A two-phase donor-acceptor bulk heterojunction lattice was generated using a simulated annealing method. Starting from a fine intermixed phase where lattice points in a cubic lattice with a lattice constant $a=$\SI{1}{\nm} are assigned randomly as donor or acceptor sites, 
cellular automata simulations~\cite{PeumansNat03,WatkinsNano05,MarshJAP07} are run to yield decreasing interfacial areas up to the point where a target domain size of \SI{8}{\nm} is reached that will characterize the phase separation length of the blend.
Although this method produces a two-phase discretized system that is only a crude approximation to the polymer:PCBM blend, it is sufficient to illustrate the effect of the relative orientation of the donor-acceptor interfaces with respect to the applied electric field. These morphologies serve as the input lattice for the Monte Carlo simulation of charge recombination. Here we use a lattice of $\SI{50}{\nm} \times \SI{50}{\nm} \times \SI{100}{\nm}$ with hard-wall boundary conditions for charges and an electric field is applied along the long axis.

Each Monte Carlo transport simulation starts by placing a number of electron-hole pairs, typically 20, at a given initial separation $L_{s}$, which allows us to start the simulation at $t=0$ with the electron-hole pair already at an intermediate separation distance. By doing that we disentangle the initial charge separation process from the long-time recombination dynamics. This is important because, firstly, tackling the dissociation problem with Monte Carlo simulations can be problematic, particularly in systems that show ultrafast charge separation~\cite{SimonScience13,SavoieJACS14}, and secondly this allows us to focus only on those charge carriers that have survived recombination from exciton-generated charge-transfer states as these are the carriers which are detected in the resonator experiment. As we will show below this assumption is critical in order to trace the long-time dynamics. The electron and hole are always generated in acceptor and donor phases respectively, and the initial vector joining the particles is always parallel to one of the lattice axes. (Relaxing this condition to allow a more realistic distribution of initial positions does not alter the qualitative behavior observed, but, due to the discrete nature of the lattice, makes it impossible to define the initial separation uniquely.) Charge hopping rates are calculated for each particle in the system according to the Miller-Abrahams formulation~\cite{MillerPR60}: 

\begin{equation} 
\nu_{ij}=\left\{ \begin{array}{ll}
 \nu_{0}\exp(-2\gamma r_{ij})\exp\left(-\frac{\epsilon_j-\epsilon_i}{k_BT}\right) &\mbox{ for $\epsilon_j>\epsilon_i$} \\
  \nu_{0}\exp(-2\gamma r_{ij}) &\mbox{ for $\epsilon_j\leq \epsilon_i$}
       \end{array} \right.
\label{eq:MArates}
\end{equation}
where $i$ denotes the residence site of the charge and $j$ the target site, and the two sites are separated by a distance $r_{ij}$. Site energies $\epsilon_i$ and $\epsilon_j$ include contributions from the static Gaussian disorder~\cite{HeinzPhysStatSol93}, the Coulomb potential due to charges in the system, and the voltage drop due to the applied field. Static disorder in the energies for the donor and acceptor phase sites has been taken from a normal distribution with a standard deviation $\sigma=\SI{0.1}{\eV}$. The inverse localisation length $\gamma=$\SI{2}{\nm^{-1}}, the attempt-to-hop frequency $\nu_0=\SI{e12}{\second^{-1}}$ and $T=\SI{10}{\kelvin}$. Using a different value for $\nu_0$, which is typically of the order of the phonon vibration frequency, would simply rescale the time evolution and we therefore show results for the time dependence using the normalized time $t/t_0$ (where $t_0$ is introduced below). It is important to note here that Miller-Abrahams type of rates are appropriate for describing hopping conduction in the non-equilibrium regime~\cite{FishchukPRB13} where the transport of particles is dispersive~\cite{HoffmannJPCC12,StavrosJPCL13,ScheidlerChemPhysLet94}. It has been shown that they can consistently reproduce the experimental temperature dependence of triplet energy relaxation and diffusivity in various polymeric and oligomeric systems~\cite{HoffmannJPCC12}. Moreover, Monte Carlo simulations with Miller-Abrahams rates~\cite{HoffmannJPCC12} agree with analytical results for the time dependence of the spectral diffusion and diffusivity~\cite{MovagharPRB86}. 

Since at low temperatures uphill energy hops are frozen, the particles will require to make long-distance phonon-assisted tunneling attempts to find accessible neighbors. We find that tunneling distances of at least \SI{4}{\nm} should be considered as restricting the neighbor space to first or second nearest neighbors alters the dynamics considerably. Therefore the results presented below include hopping events at distances $r_{ij}\leq \SI{4}{\nm}$. Here, in order to take into account the change in the electrostatic potential with each hopping event that will in turn influence the transfer rates, we choose a direct rejection-free kinetic Monte Carlo scheme where the potential is recalculated at every time step. Partial sums for each event are placed in an array, $s(n)=\sum_{l=1}^{n}\nu_{l}$  where $n=1,..., N$ with $N$ the total number of events and $\nu_{tot}=s(N)$ the total escape rate that takes the system from one state to the next. A random number $\chi$ from a box distribution between 0 and 1 is drawn and its product with $\nu_{tot}$ determines which event will follow according to $s(k-1)<\chi \nu_{tot}\le s(k)$. The event $k$ is chosen and the system is advanced to the new configuration with a time step  $\tau=-\frac{1}{\nu_{tot}}\ln X$, 
with $X$ a new random number, $X \in (0,1)$. The system is allowed to evolve until a termination time of $t_{term}=10^5 t_0$ with $t_0$ the minimum jump time between nearest-neighbor sites, $t_0=(1/\nu_0)\exp(2\gamma a)$. Available microscopic events are hopping events for electrons and holes and recombination events between opposite carrier types. Recombination events are included when carriers of opposite polarity are nearest neighbors and we have considered the two extreme values for the recombination rate, immediate recombination with $\nu_{rec}=\infty$ and no recombination with $\nu_{rec}=0$. One thousand starting configurations are considered and for each of these we monitor the total number of carriers in the system. Therefore averaging over all configurations, we can construct the time evolution of the survival probability
\begin{equation} 
P_{surv}(t)=\frac{N(t)}{N(t=0)},
\end{equation} defined as the average number of carriers, $N(t)$, that have survived recombination at time $t$, normalized by the starting number of carriers $N(t=0)$. In what follows, we  show results for $\nu_{rec}=\infty$. 

\begin{figure}
\centerline{\includegraphics[clip=true,width=9.5cm]{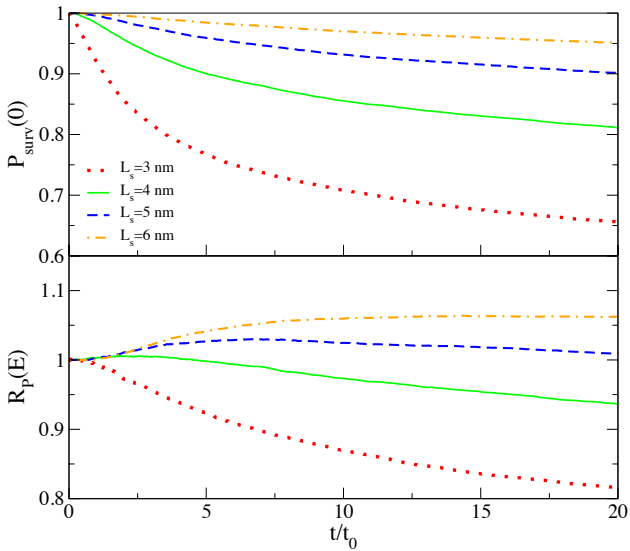}}
\caption{(Colour online) Monte Carlo simulations for a blend morphology. Top panel: carrier survival probability for electron-hole pairs generated at random interfaces with an initial separation $L_{s}$ for zero electric field. Bottom panel: ratio of the survival probabilities at zero  and high  $E=\SI{e8}{\volt/\m}$  field for different $L_{s}$.
}
\label{SurvivalTheo}
\end{figure}

The central simulation result is summarized in Figure~\ref{SurvivalTheo} and shows that although for short initial electron-hole separation $L_s$ an external applied electric field  reduces recombination, there exists a threshold value of separation above which the field can increase recombination. The top panel of Figure~\ref{SurvivalTheo} shows the time evolution of the survival probability for different values of $L_s$ at zero electric field. As expected, the survival probability decays more slowly at higher initial separations $L_s$. The bottom panel of Figure~\ref{SurvivalTheo} shows the ratio of the survival probabilities at zero and high field,
\begin{equation} 
R_{P}(E)=\frac{P_{surv}(0)}{P_{surv}(E)} .
\end{equation} This quantity is smaller than unity when the external field reduces recombination and larger than unity when the field assists recombination. It can be seen that when $L_s>$\SI{4}{\nm} recombination increases in the presence of a field. This behavior is unexpected as one might expect that in a PV blend with a low concentration of carrier pairs, an external field would assist dissociation by helping carrier pairs to overcome their Coulomb attraction. We will show below that the observed effect is related to morphology and energetic disorder.

\begin{figure}
  \centerline{\includegraphics[clip=true,width=9.5cm]{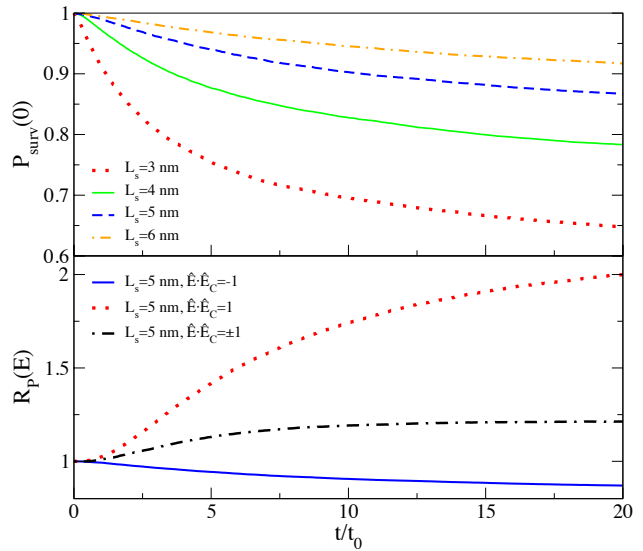}}
\caption{(Colour online) Monte Carlo simulations for a bilayer morphology. Top panel: carrier survival probability for electron-hole pairs generated at the heterojunction with an initial separation $L_{s}$ for zero field. Bottom panel: ratio of the survival probabilities at zero and high $E= \SI{e8}{\volt/\m}$  field for $L_{s}=$ \SI{5}{\nm} and different field orientations: $\vec{E}$ against the e-h pair Coulomb field $\vec{E}_{C}$ (solid line), $\vec{E}$ along $\vec{E}_{C}$ (dotted line) and $\vec{E}$ either along or against $\vec{E}_{C}$ with equal probability (dashed-dotted line).}
\label{SurvivalBilayer}
\end{figure}

We can understand this behavior by considering the effects of the electric field on carrier pairs oriented in different directions with respect to the applied field, such that the applied field either assists or opposes the Coulomb attraction between the electron and holes.  To illustrate this we perform Monte Carlo simulations in bilayer geometries, where the electric field and the initial pair separation are perpendicular to the plane of the donor interface. The top panel of Figure \ref{SurvivalBilayer} shows the time dependence of the survival probability for different initial e-h separation distances in the absence of a field. As expected, and similar to the blend, recombination is slower for larger initial separations.  The bottom panel of Figure~\ref{SurvivalBilayer} shows ratio of survival probabilities at zero and high field  $R_{P}(E)$ for different field directions, for an initial separation of \SI{5}{\nm}.  As expected, when the electric field acts against the e-h pair Coulomb field (case (i)) it reduces recombination. When the applied field is along the Coulomb field (case (ii)), recombination is enhanced.  This enhancement in $R_{P}(E)$ is more pronounced than the reduction seen in the previous case.  As a simple approximation to the situation in a blend, we also consider a simulation where for half the trials the configuration is in case (i) and for the other half it is in case (ii).  Here recombination increases with applied field; the case (ii) behavior dominates over the case (i) behavior, reproducing the same effect seen in the blend.  We note that the field-induced recombination enhancement only occurs for initial separations greater than \SI{4}{\nm}, and for large values of $\sigma / k_{B}T$.

The simulations presented so far assume infinitely fast recombination when an electron and a hole reach a minimum separation distance, but we have also considered  the case where recombination rate is zero. In that case similar conclusions are reached if instead of the survival probability we monitor the time and field dependence of the number of e-h pair encounters. Thus the long time-dynamics of the recombination process are not reaction-limited but rather transport-limited.

 Given that here we monitor the decay of an initial e-h population generated at $t=0$, the magnitude and time dependence of the quantity $R_{P}(E)$ is sensitive to the initial distribution of carriers.  In fact, for $L_{s} \ge \SI{4}{\nm}$, $R_{P}(E)$ will reach a peak and will start decreasing when most of the e-h pairs vulnerable to recombination with Coulomb field $\vec{E}_{C}$ along $\vec{E}$ have recombined. When this occurs, the average e-h pair distance is larger than the initial $L_{s}$ due to surviving e-h pairs with $\vec{E}_{C}$ against $\vec{E}$ being further separated. Therefore, $P_{surv}(E)$ will decay slower than $P_{surv}(0)$ and $R_{P}(E)$ will decrease with further increases in the simulation time, resulting in a crossover time where $R_{P}(E)$ becomes smaller than unity. Clearly this crossover time will be longer for larger initial e-h separation distances and within the time window presented in Figure \ref{SurvivalTheo} it is reached only for $L_s=$\SI{4}{\nm}. In the experiments presented here this regime is not accessible because of the continuous illumination conditions.

\section{Discussion}
The experimental and simulation results above clearly show that at low temperatures in a bulk heterojunction polymer:PCBM blend the overall influence of an external electric field on the long-time dynamics of photogenerated carriers is to enhance recombination. This contrasts with the commonly held picture, arising from an effective medium model, where the field is assumed always to assist in separating carriers. 

\begin{figure}
  \centerline{\includegraphics[clip=true,width=9cm]{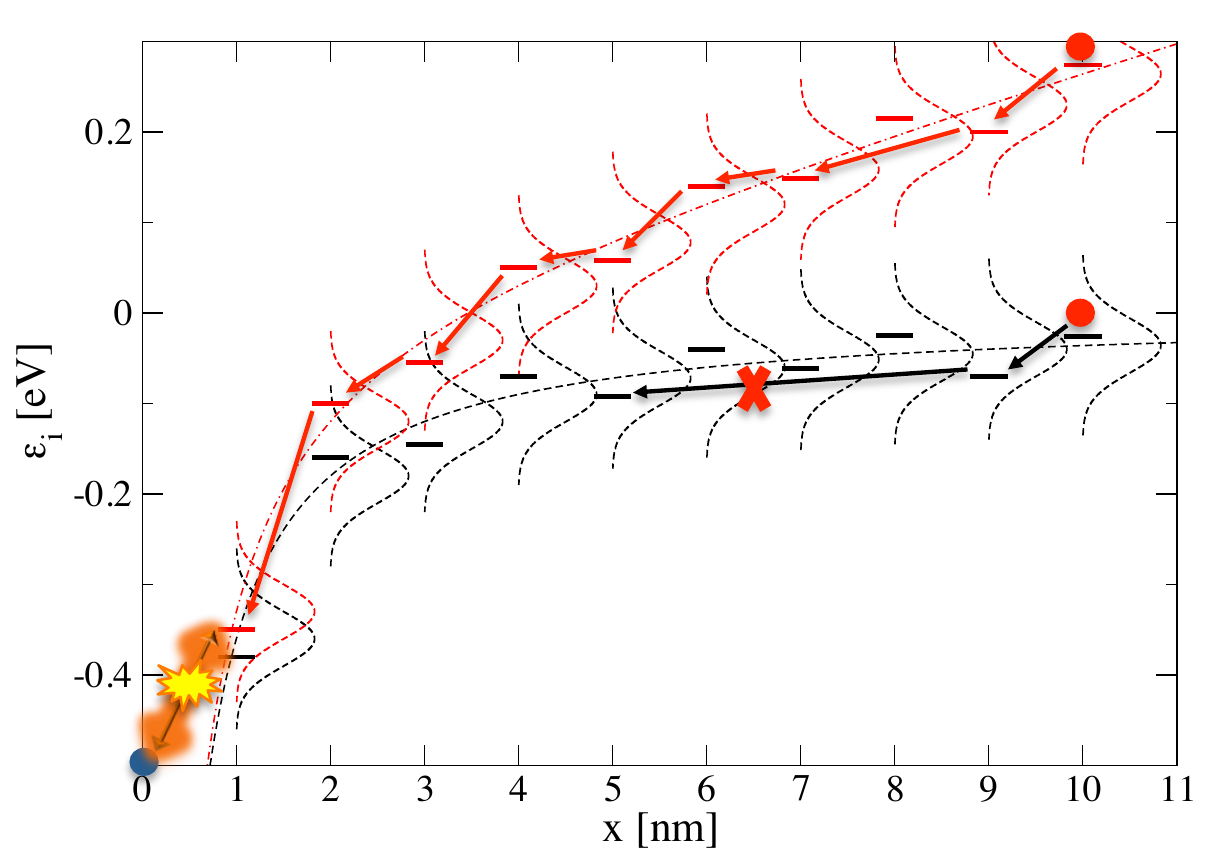}}
\caption{(Colour online) Pictorial representation of the influence of an electric field on low-temperature recombination. Electron site energies follow the Coulomb potential (black dashed line) generated by a hole placed at $x=0$. An external electric field $\vec E\uparrow\uparrow \vec E_C$ will modify the electron energy sites (red dashed-dotted line). In the absence of disorder, for an electron placed at a large distance (here shown $x=$\SI{10}{\nm}) the potential landscape will be downhill and recombination will occur when the electron reaches next to the hole at $x=$\SI{1}{\nm}.  Disorder on the energy sites, shown by the Gaussian dashed curves, will create energy barriers (example site energies shown by segments) on the route to recombination and for zero field long-distance tunneling events are required for reaching the interface, preventing recombination.  $\vec E$ will decrease the number of long-distance jumps and increase the probability for recombination.
}
\label{pictorial}
\end{figure}

To understand in more detail the microscopic origin of this effect, we consider the situation shown schematically in Fig.~\ref{pictorial}  where at $t=0$ an electron and a hole are separated by distance $L_s$. For simplicity we place the hole at $x=0$ next to the interface, where the electron has to travel in order to recombine. The energies of the electron hopping sites are modified according to the Coulomb potential generated by the hole that scales as $1/x$. At low temperatures, in the absence of any external field and for zero disorder the electron will experience a potential landscape that is always downhill towards the hole.  The electron will therefore move towards the hole with short-distance jumps; in the Miller-Abrahams model these jumps will be of a constant rate. Now let us switch on Gaussian site energy disorder. The situation now changes as the energy landscape roughens and the electron is surrounded by sites that can be of higher or lower energy. In fact, the further away the electron is from the interface, the more likely it is to get blocked en route as the Coulomb interaction is not sufficient to overcome energy barriers, and in the absence of any activation mechanisms the probability of electron-hole encounter is zero. An external electric field will provide such an activation channel and it will allow the electron to access sites closer to the interface where the Coulomb interaction with the hole is sufficient to overcome barriers due to Gaussian energetic disorder. If the direction of the field is against the e-h pair Coulomb field, $\vec E \uparrow\downarrow \vec E_C$, then on average it will assist on separating the charges, but these charges would otherwise not recombine either, therefore the probability for recombination remains almost zero. Note that for a three-dimensional morphology there is a small possibility that such a field will allow the electron to find another path that will lead to recombination. If the field has a component along the Coulomb field, $\vec E\uparrow\uparrow \vec E_C$, then it will allow sites closer to the hole to be accessed, and will increase the probability for recombination. In a bulk heterojunction blend, switching on an external field will allow the pairs for which $\vec E\uparrow\uparrow \vec E_C$ to recombine whereas the e-h pairs with $\vec E \uparrow\downarrow \vec E_C$ will separate further.

\section{Conclusions}
In conclusion, we have developed a new microwave resonator technique to monitor charge recombination in organic photovoltaic blends. This method allows us to assess the influence of an external electric field on the population of the carriers and the rate of recombination at cryogenic temperatures. In contrast to the normal case, where electric field decreases recombination, we find that recombination can be dramatically enhanced by the application of a static electric field. A Monte Carlo hopping model that includes tunneling events has been employed to understand at the microscopic level the field-dependent recombination under non-equilibrium conditions. This model shows that morphology plays a crucial role in determining the dynamics of recombination, which under certain conditions can be enhanced for heterojunction donor:acceptor type of systems with nanoscale phase separation. This is a subtle effect and comes as a consequence of the interplay between charge separation distances, energetic disorder and interfaces that are against the field assisting charge extraction. In a bulk heterojunction PV blend these conditions are met when the electron-hole separation is larger than \SI{4}{\nm}. The above scenario is consistent with recent transient absorption experiments~\cite{SimonScience13} that suggest ultrafast separation of charge-transfer states in polymer:fullerene blends on a several nanometer length-scale within quasi-crystalline fullerene domains. Very recently, e-h distance distributions for separated charges that peak at $\sim$ 3-\SI{4}{\nm} have been inferred~\cite{HodgkissJACS14} from transient absorption spectroscopy at low temperatures for a range of polymer:PCBM blends.

\acknowledgments This work is supported by the UK Engineering and Physical Sciences Research Council (grant number EP/G060738/1).
We thank H. Bouchiat for fruitful discussions and support in the fabrication of the resonators and S. Bayliss for carefully reading the manuscript and providing comments. 
A.C. acknowledges support from the E. Oppenheimer Foundation and St Catharine's College, Cambridge. 
Part of the Monte Carlo simulations were performed using the Darwin Supercomputer of the University of Cambridge High Performance Computing Service (http://www.hpc.cam.ac.uk/), provided by Dell Inc. using Strategic Research Infrastructure Funding from the Higher Education Funding Council for England and funding from the Science and Technology Facilities Council.

\bibliography{PRX_paper_v3final}

\end{document}